\documentstyle[11pt,newpasp,twoside,epsf]{article}
\def\edcomment#1{\iffalse\marginpar{\raggedright\sl#1\/}\else\relax\fi}
\reversemarginpar

\begin{document}
\title{Recurrent Novae, Classical Novae, Symbiotic Novae, and Population II
      Novae} 

\author{Mariko Kato} 

\affil{Keio University, 4-1-1, Hiyoshi, Kouhoku-ku, Yokohama, 223-8521, Japan}

\begin{abstract}
Light curve analysis of decay phase of nova outburst are summarized.
Nova duration is determined by the strong wind mass-loss which depends 
only on the white dwarf mass. Fast novae correspond very massive white
dwarfs and very slow novae correspond almost lower mass limit 
of white dwarfs.
This relation is shown in both of classical novae and recurrent novae. 
Light curves are theoretically reproduced for recurrent novae,
U Sco, V394 CrA, CI Aql, V2487 Oph, RS Oph, T CrB,  V745 Sco and  V3890 Sgr,   
using binary model composed of a white dwarf, an irradiated accretion disk, 
and partly irradiated companion which is shadowed by the accretion disk.
From the light curve fitting, we conclude that most of these objects 
contain a very
massive white dwarf close to the Chandrasekhar
mass limit (1.37 $M_{\odot}$ for U Sco, V394 CrA T CrB and RS Oph,
1.35 $M_{\odot}$ for V2487 Oph, V3890 Sgr and V745 Sco). They are
strong candidates of type Ia SN progenitors. 
Population II novae have trends of slow evolution and small expanding 
velocity compared with disk novae. 
\end{abstract}

%%%%%%%%%%%%%%%%   section 1   %%%%%%%%%%%%%%%%%%%
\section{Nova Outbursts and Optically Thick Wind Theory}
Nova outbursts are thermonuclear runaway events on white dwarfs.
Depending on the binary parameter such as the mass accretion rate, 
composition of the
envelope and the white dwarf mass, nova outbursts shows a wide variety of
the recurrence period, its light curve, the duration time and expanding 
speed of gaseous matter and so on. 

Classical nova shows enhancement of carbon and oxygen or other white 
dwarf material in its ejecta. The recurrence period is as long as ten thousand
of years or so and we can observe only one outburst for one nova binary. 
On the other hand, recurrent novae repeat outbursts every 
several to several-tens of years and its ejecta shows no enhancement of 
white dwarf material. 

After the onset of shell flash, the white dwarf rapidly brightens 
up and the envelope greatly extend to a giant size and the strong wind 
mass-loss begins to blow the envelope mass.
After the star reaches the optical maximum the photospheric radius reduces
with increasing effective temperature, which causes the decay of 
visual light curve. 

The mass loss during nova outburst is a radiation-driven wind  
which is accelerated deep inside the photosphere (Friedjung 1966).
Such a wind is called as the optically thick wind. The decay phase of 
nova outbursts is followed by  a quasi-evolution theory (summarized in 
Kato \& Hachisu 1994) which is to make a sequence of optically thick 
wind solutions of expanding envelope. 

%%%% figures 1 (kato_fig1.eps, kato_fig2.eps opacity and structure)
\begin{figure}
%\plottwo{kato.opac.eps}{kato.vt.eps}
\plottwo{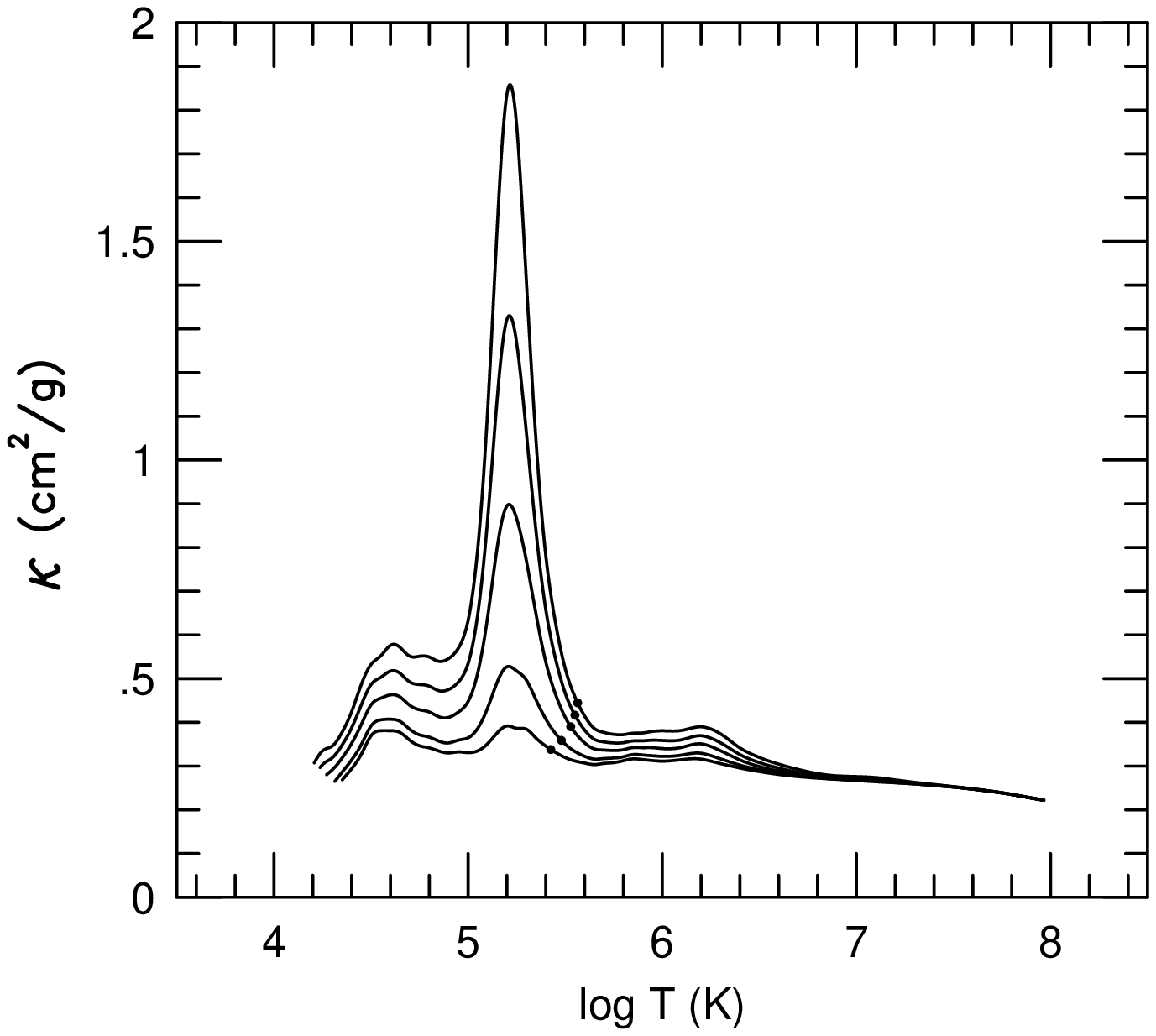}{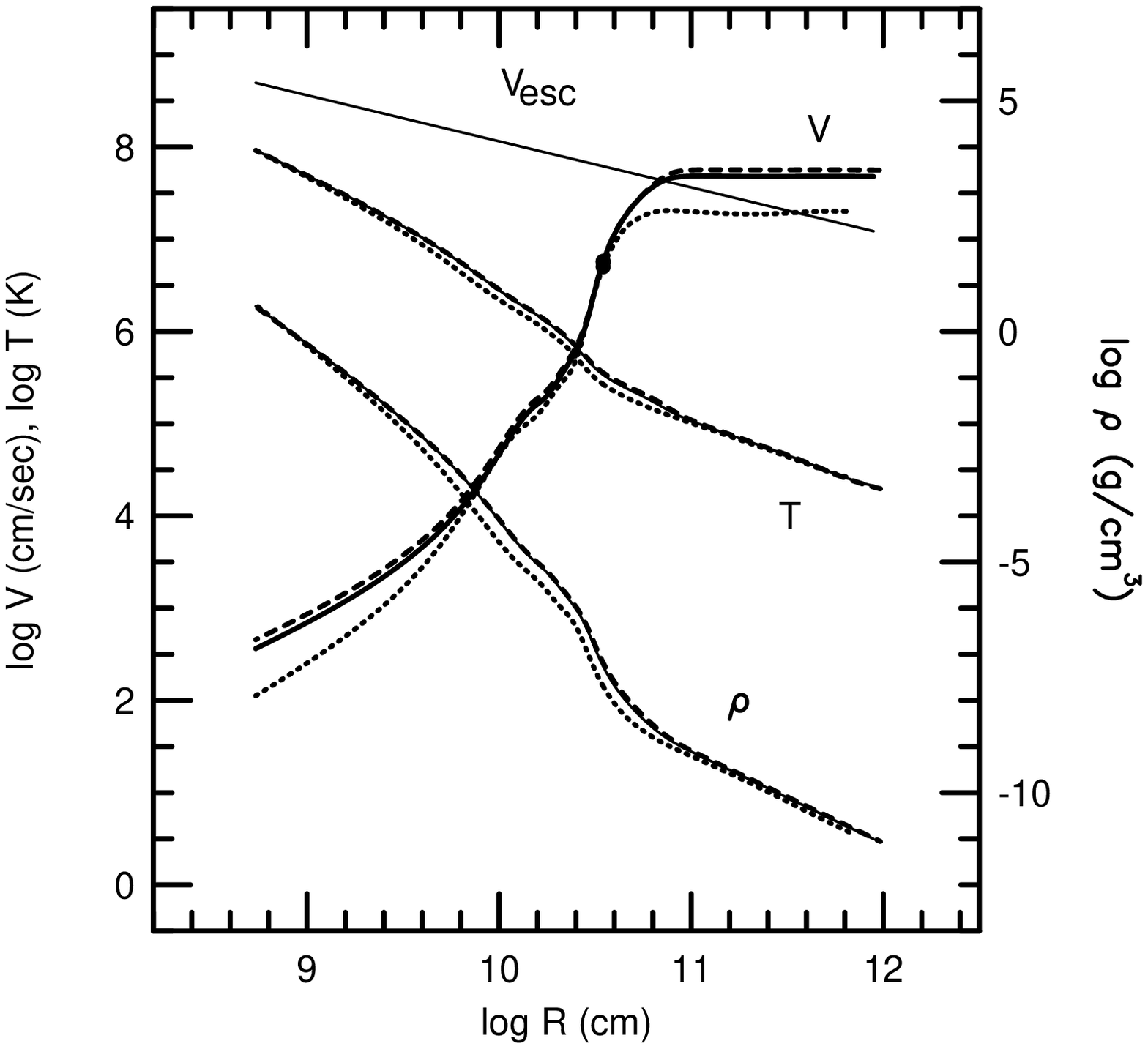}
\caption{(left) Run of the updated OPAL opacity for the wind solutions
of a classical nova model on a $1.0 M_{\odot}$ white dwarf. The heavy 
element contents is $Z=0.1, 0.05,0.02, 0.004$ and $0.001$ from top to bottom.
For the detail, see Kato 1997.
(light) Change of the velocity ,the temperature, and the density of three
envelope of a classical nova model. $Z=0.05$ (dashed line), 0.02(solid line),
and 0.001 (dotted line).
}
\end{figure}

Figure 1 (left) shows 
the run of the opacity throughout an envelope of a classical
nova model. The OPAL opacity has a strong peak owing to iron lines at
$\log T \sim 5.2$ which locates deep inside the photosphere. This opacity peak
blocks radiative flux to accelerate the wind mass-loss.  As shown in the right
figure 
the wind velocity increases quickly at the temperature region corresponding 
to this opacity peak. The wind velocity already reaches the terminal 
velocity at the photosphere. In the optically thick wind the mass-loss rate 
is very large, because the acceleration occurs deep inside where the density
is high. It is naturally understood that the velocity and the mass-loss rate 
of winds are larger for larger metallicity, i.e., larger in population I 
than in population II novae.

%%%%%%%%%%%%%%%%   section 2   %%%%%%%%%%%%%%%%%%%
\section{The HR diagram and the Decay Time-Scale}
%%%%figure 2
\begin{figure}
%\plotone{kato.hr.eps}
\plotone{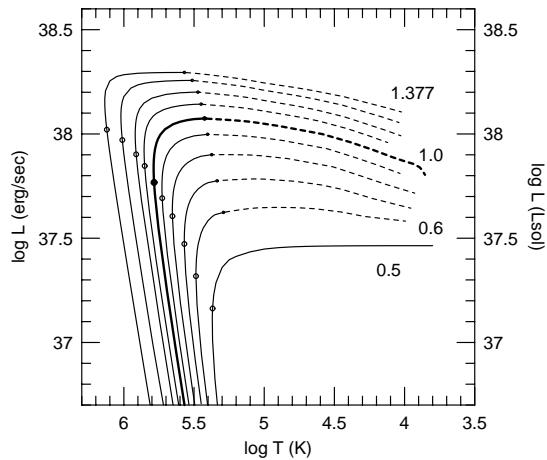}
\caption{Theoretical H-R diagram for the decay phase of recurrent nova.
The white dwarf mass is $1.377, 1.3, 1.2, 1.1, 1.0, 0.9, 0.8, 0.7, 0.6,$ 
and $0.5 M_{\odot}$. The chemical composition of $X=0.7$ and $Z=0.02$ 
is assumed. The optically thick wind occurs in the dashed region.  
Hydrogen burning extinguishes at the point marked by the filled circle.}
\end{figure}

The evolutionary course of the decay phase in H-R diagram is plotted in 
figure 2. The decay phase of the classical nova and the recurrent nova
with different composition is similar to the curves in figure 2 (see 
Kato 1997 for classical novae,  Kato 1999 for recurrent novae).

At the optical peak the star reaches the maximum radius which point locates 
somewhere in the right side of each curve. 
As the strong wind blows out the envelope mass, the photospheric 
radius moves inward and the effective temperature rises, thus the star moves
leftward. The wind mass-loss occurs in the dashed parts. After the
wind stops, the star still goes leftward until nuclear burning extinguishes
at the small dot. In less massive stars, the acceleration is 
not enough to cause the optically thick wind.  

In massive white dwarfs, the evolutionary speed is short mainly because of 
the small envelope
mass which is blown off within a short time by the wind. 
Therefore, the nova outburst is fast in massive white dwarfs and 
slow in less massive stars. In a lowest limit of white
dwarfs, the acceleration is weak to cause the optically thick wind. 

The wind acceleration also depends on the metallicity as shown in the 
left figure. The larger iron content, the stronger peak of the opacity 
that causes a strong wind. 
In $Z = 0.02$, shell flashes do not occur or very weak in $\leq 0.4 
M_{\odot}$ white dwarf for classical novae, and  $\leq 0.5 M_{\odot}$ for 
recurrent/symbiotic novae.
In $Z \leq 0.004$, shell flashes do not occur or very weak in $\leq 0.6 
M_{\odot}$ for classical nova, and  $\leq 0.8 M_{\odot}$ for 
recurrent/symbiotic nova.

%%%%%%%%%%%%%%%%   section 2   %%%%%%%%%%%%%%%%%%%
\section{Light Curve Analysis  and Decay Time-Scale}

\subsection{Classical Nova}
The decay phase of nova outburst is followed by the optically thick wind
theory. The method of light curve fitting of nova is firstly established 
in nova Cygni 1978 (Kato 1994).  The wind mass-loss rate depends strongly 
on the white dwarf mass, therefore, we can determine the white dwarf mass 
from the light curve fitting. In case of nova Cygni 1978, the two independent 
fittings of optical and UV light curves give the same values of white dwarf 
mass of $1.0 M_{\odot}$ and the distance to the star. 

Nova Cygni 1978 is a moderately fast nova. 
More rapid evolution is predicted only in more massive white dwarfs. 
Light curve fittings of several 
classical novae show a fast classical nova corresponds to a very massive 
white dwarf ($\sim 1.3 M_{\odot}$), moderately classical nova such as Nova 
Cyg 1978, correspond a moderately massive white dwarf
($\sim 1.0 M_{\odot}$), and slow evolution of slow novae, the less 
massive white dwarf($\sim 0.6 M_{\odot}$ for Nova Mus 1985). It is very 
interesting that the fastest limit of classical nova corresponds to the 
massive limit of the white dwarf , and the slowest one the lowest limit.

\subsection{Recurrent Novae and Symbiotic Novae}
Recurrent nova and symbiotic nova can be treated as a same subgroup 
of nova, that shows no white dwarf material enhancement. The difference 
between the two is in the white dwarf mass and the contribution of the companion
and the accretion disk. The similar relation to classical nova,
i.e., the relation between the evolutionary speed 
of light curves and white dwarf mass, also stands 
in these novae.  
Most rapid evolution of some recurrent novae corresponds very massive
white dwarf close to the upper limit ($\sim 1.37 M_{\odot}$ as shown above) 
and slow evolution  the moderately massive white dwarfs (e.g., $\sim 1.2 
M_{\odot}$ for T Pyx). Very long duration of some symbiotic novae corresponds
low-mass white dwarfs ($ \sim 0.6 M_{\odot}$ for symbiotic nova RX Pup).

%%%%%%%%%%%%%%%%%%%%%    Recurrent Nova   %%%%%%%%%%%%%%%%%%%%%%%%%%%%%%%%%
\section{Light Curve Analysis of Recurrent Novae}
Using binary models composed by a white dwarf, an accretion disk and a
companion,
we have calculated theoretical light curve of recurrent novae. 
In the first phase, the envelope largely expand to about several tens of the
binary size, and the companion are deeply embedded by the 
envelope. As the strong wind carries away most of the expanding 
envelope, the photospheric radius moves inward, and in the later phase, 
we can see the companion and the accretion disk again. 

The early phase of the light curve is determined by the 
wind mass-loss rate. In the following plateau phase the accretion disk and
the partly irradiated companion contribute the visual magnitude. Therefore, 
in this stage, we have calculated visual flux summing up of three component, 
of the white dwarf photosphere, the irradiated accretion 
disk, and the partly irradiated companion that is shadowed by the 
accretion disk. We divided their surfaces into 
9000 small areas and calculated emission flux from each areas to sum up. 

\subsection{U Sco type subclass}
The recurrent nova U Sco, V394 CrA, CI Aql, and V2487 Oph are members 
of the subclass of U Sco type that have a slightly-evolved main-sequence 
companion.

U Scorpii is one of the well observed recurrent novae, characterized by 
the shortest recurrence period $\sim 8$ yr, the fastest decline of its 
light curve 0.6 mag per day, its extremely helium-rich ejecta
He/H$\sim 2$ by number. The  latest outburst in 1999  was well observed
from the rising phase to the cooling phase by many observers including 
eclipses which provide us a unique opportunity 
to construct a comprehensive model for U Sco during the outburst.

We have reproduced the light curves by binary model as shown in figure 3.
To fit the early linear decay phase ($t \sim 1-10$ days after 
maximum), we have calculated total 140 $V$-magnitude light curves with
various WD mass, hydrogen content the envelope, where the heavy element 
content $Z=0.02$ is fixed, and companion mass.  We choose $1.377 M_\odot$
as a limiting mass just before the SN Ia explosion in W7 model ($M_{\rm Ia}
=1.378 M_\odot$) of Nomoto, Thielemann, \& Yokoi (1984).

The early 7 days light curve hardly depends on the chemical composition 
or the companion mass but mainly depends on the white dwarf mass.
The $1.37 M_\odot$ light curve is in much better agreement with the 
observations than the other WD masses.
In the middle plateau phase, the irradiated accretion disk mainly 
contributed to $V$ light, because the white dwarf photosphere shrinks to
a small size and shadowed by the accretion disk.

To fit the cooling phase ($t \sim 30-40$ days after maximum),
we must adopt very small  hydrogen content of $X=0.05$
among other values. 
This is because the hydrogen content, $X$, determines the period of 
hydrogen steady shell burning, i.e., the period of the mid-plateau phase.
In our model, the optically thick wind stops at $t= 17.5$ days, and
the steady hydrogen shell-burning ends at $t= 18.2$ days.

In the evolutional course of the outburst, the visual magnitude drops 
and  UV and soft X-ray fluxes increases. It is because the photospheric 
temperature increases after the optical peak, because the main emitting region 
moves blueward (to UV then to soft X-ray).
This picture is very consistent with the BeppoSAX supersoft X-ray detection 
$19-20$ days after the optical peak (Kahabka, et al. 1999). 
In the strong wind phase we cannot expect much of supersoft X-rays because of 
self-absorption by the wind itself. 

In this way, the light curve fitting gives us an estimate for binary 
parameters, such as the white dwarf mass and  chemical composition of the 
envelope, and if orbital period is available, we further know the mass 
transfer rate and the growth rate of the white dwarf mass. 

%%%%%%%%%%%%%%%%    figure 3  %%%%%%%%%%%%%%%%%%%%%%%%%%%%%%
\begin{figure}
%\plottwo{kato.usco1.eps}{kato.usco2.eps}
\plottwo{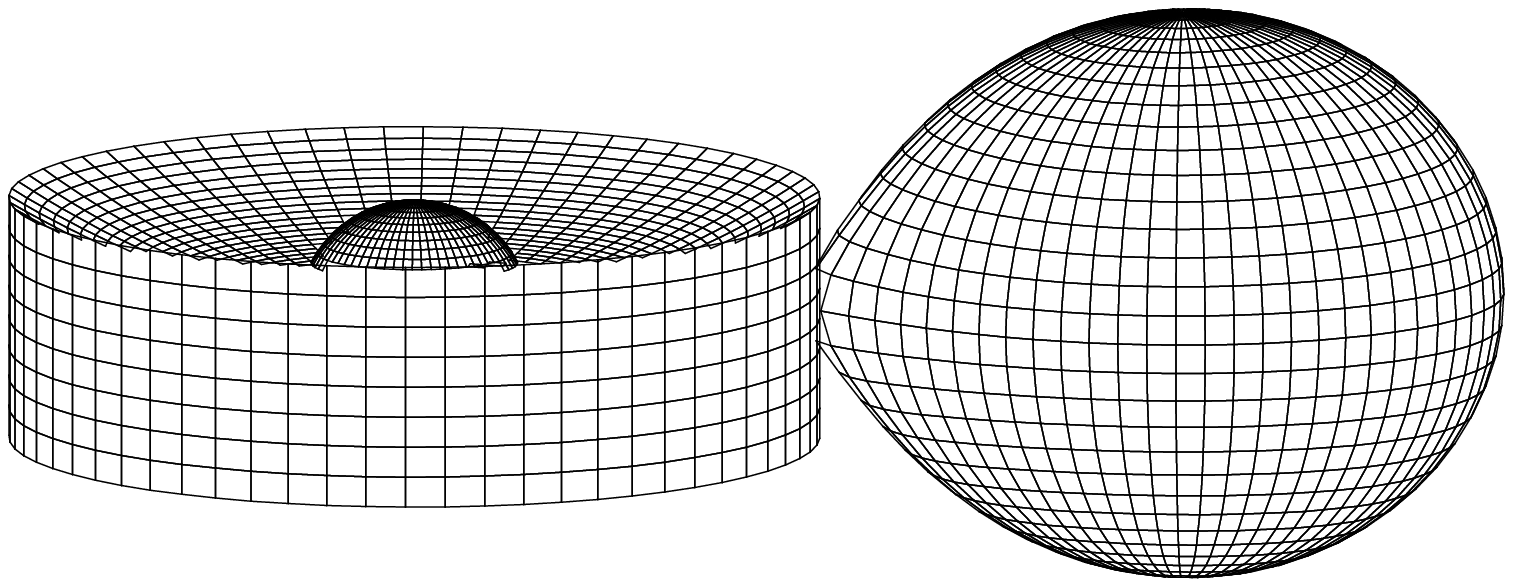}{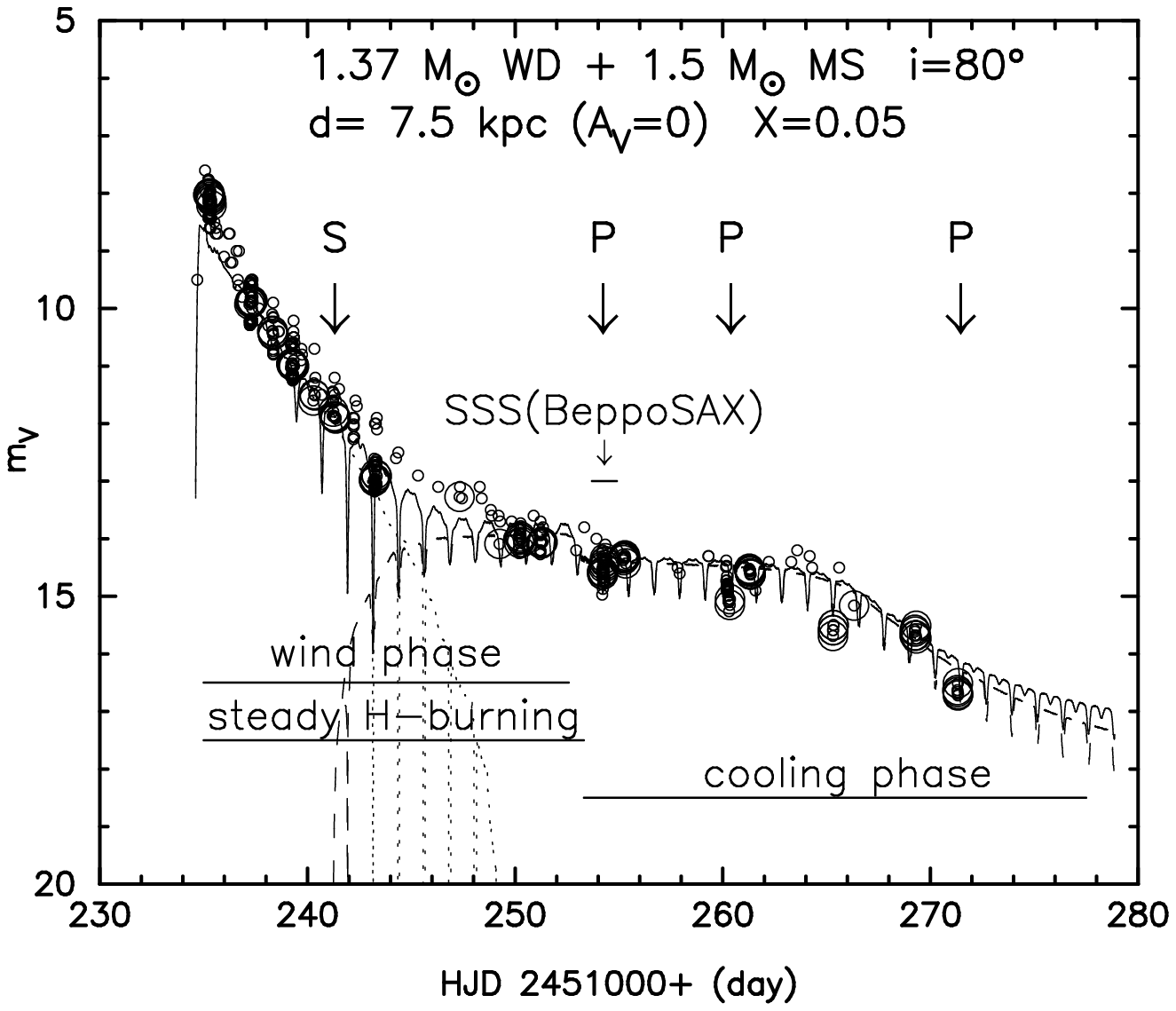}
\caption{(left) Configuration of our U Sco binary model. About 9 days
after the maximum, the white dwarf photosphere shrinks to $R_{\rm ph} = 1.0
R_{\odot}$. (light) Theoretical light curve (solid line) and observational
points. The contributions of the white dwarf photosphere and the accretion 
disk which is partly irradiated is denoted by dotted and dashed curves,
respectively. For details, see Hachisu et al. 2000.
}
\end{figure}

V394 CrA is a twin system that shows very similar light curves. 
In the same way as in U Sco, we have reproduced light curve of V394 CrA and
determined the white dwarf mass to be 1.37 $M_{\odot}$  and $X=0.05$ with
$Z=0.02$ as shown in figure 4 (Hachisu \& Kato 2000b)

\begin{figure}
%\plottwo{kato.v394cra1.eps}{kato.v394cra2.eps}
\plottwo{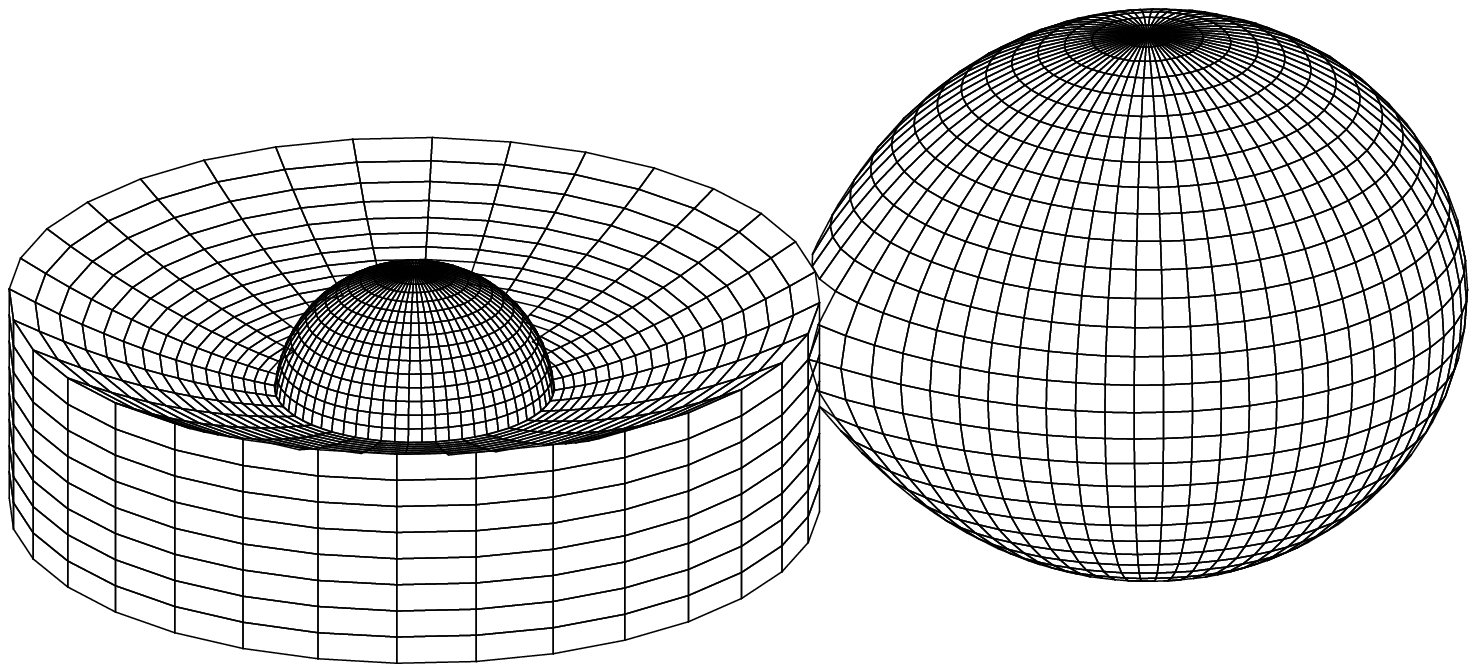}{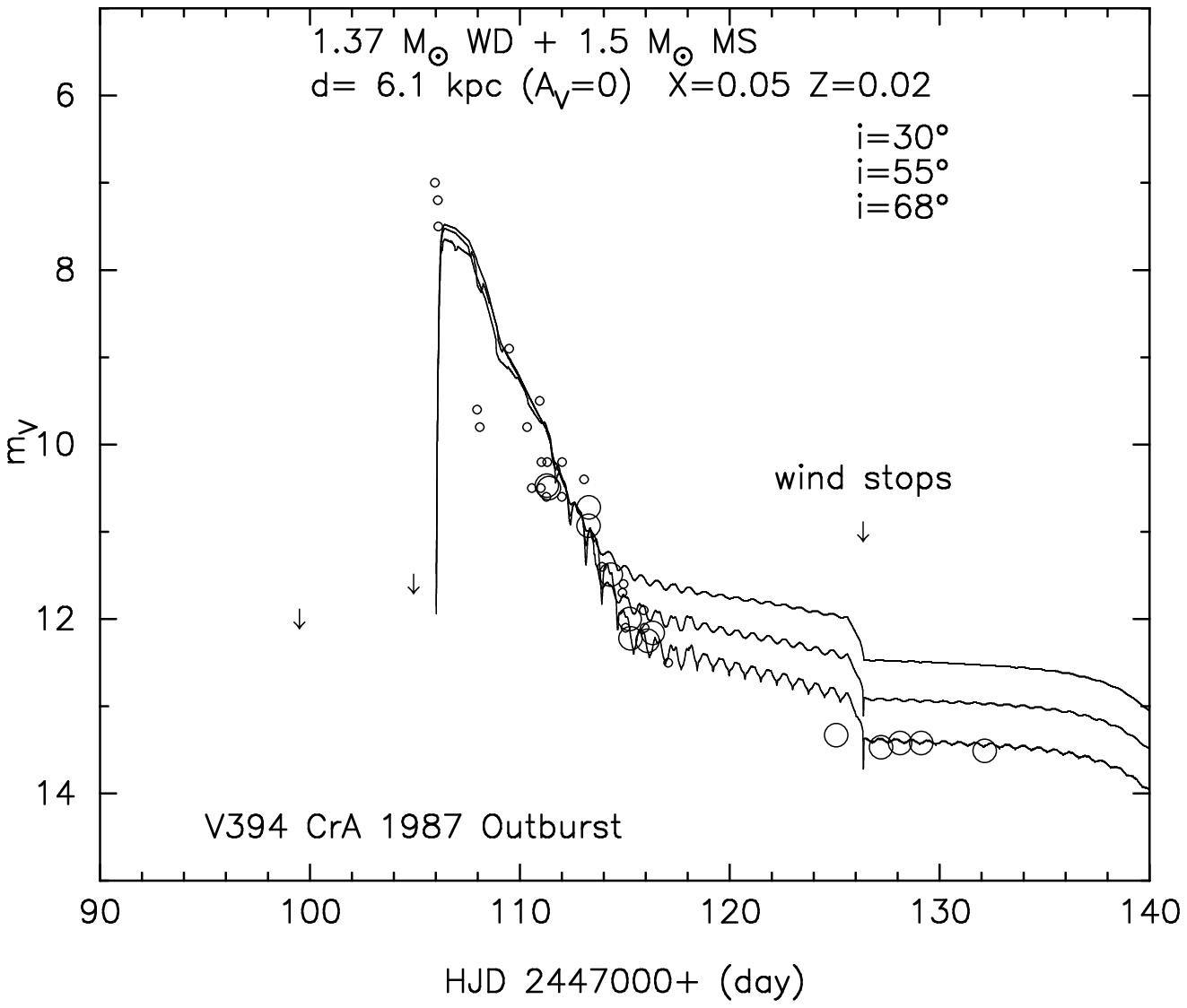}
   \caption{(left) Binary model of V394CrA about 11 days after the maximum.
   the white dwarf photosphere shrinks to $R_{\rm ph} = 1.0 R_{\odot}$. 
(light) Theoretical light curve for V394CrA (solid lines) with three 
inclination angles (${\it i}=30 \deg$ for the top curve). Observational
points are indicated by circles. For details, see Hachisu \& Kato 2000b.
}
\end{figure}

%xxxx
The light curve analysis of CI Aql and V2487 Oph is separately reported 
in detail in this conference proceedings (Hachisu et al. for V2487 Oph, 
Hachisu \& Kato for CI Aql). The white dwarf mass of these two recurrent 
novae are also massive, 1.2 $M_{\odot}$ for CI Aql and 1.35 $M_{\odot}$ 
for V2487 Oph.

\subsection{RS Oph type subclass}
RS Oph, T CrB,  V745 Sco and  V3890 Sgr are a subclass of recurrent nova
with a red giant companion. As the companion substantially contributes 
the light curve in later phase, we have calculated huge number of binary models
with different parameters and choose a configuration which gives the best 
fitted light curve.

Figure 5 (left) shows the binary configuration of T CrB selected in 
this way. The cool component (right) is a red giant filling up its inner 
critical Roche lobe. Its hemisphere is heated up by the hot component
($\sim 1.37  M_{\odot}$ white dwarf, left). The surface of the accretion 
disk is also heated up. 

The light curve fitting is shown in the right figure. 
If we do not include the contribution of irradiation from the companion, 
the summation of the $V$ light of the white dwarf photosphere and the red 
giant photosphere is very small in the later phase as shown by the dotted 
line. The dash-dotted line denotes the case if we include the irradiation 
from partly heated RG photosphere, and the solid line the case if we further
include the irradiation form the  accretion disk.

We assume a tilting accretion disk as in the left figure with precession 
at a period of about 140 days. The disk in T CrB meets the condition of 
instability against radiation-induced warping (Pringle 1996) with the
disk size, bolometric luminosity, the mass of the white dwarf 
mass accretion rate obtained our light curve fitting.
The accretion disks in RS Oph and other two systems do not meet this 
instability condition mainly because of small disk size. Therefore, we 
cannot expect the secondary peak for these other three systems. 
This is the reason that makes T CrB a very unique nova with the prominent 
secondary peak (Hachisu \& Kato 2001).

%%%%%%%%%%%%%%%%  figure 5 for  T CrB  %%%%%%%%%%%%%%%%%%%%
\begin{figure}
%\plottwo{kato.tcrbb.eps}{kato.tcrb.eps}
\plottwo{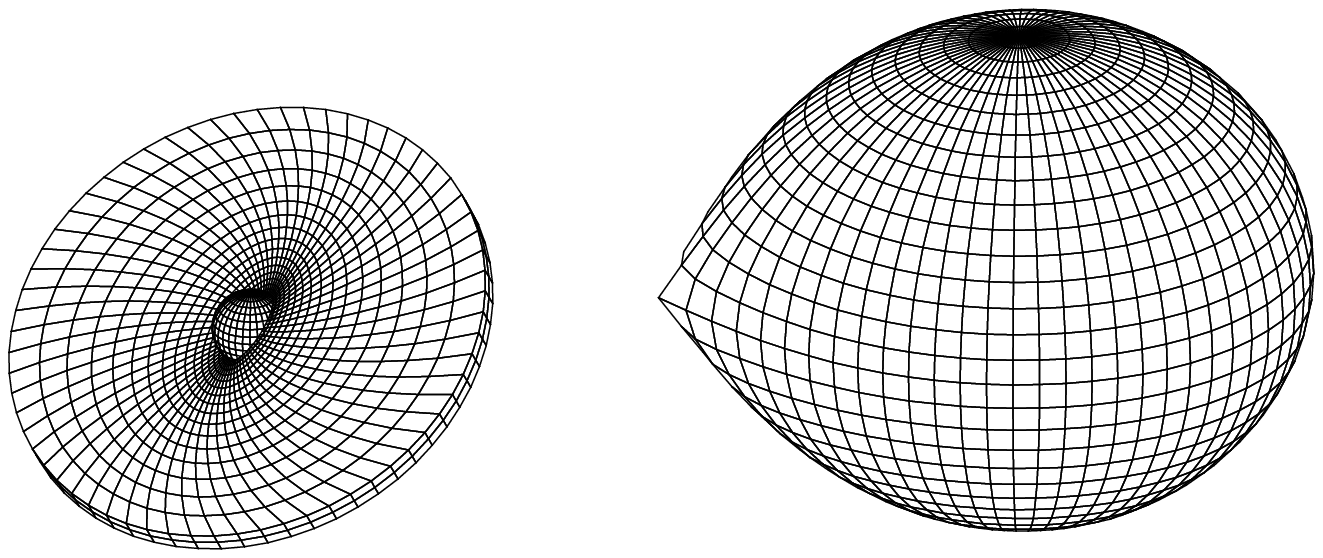}{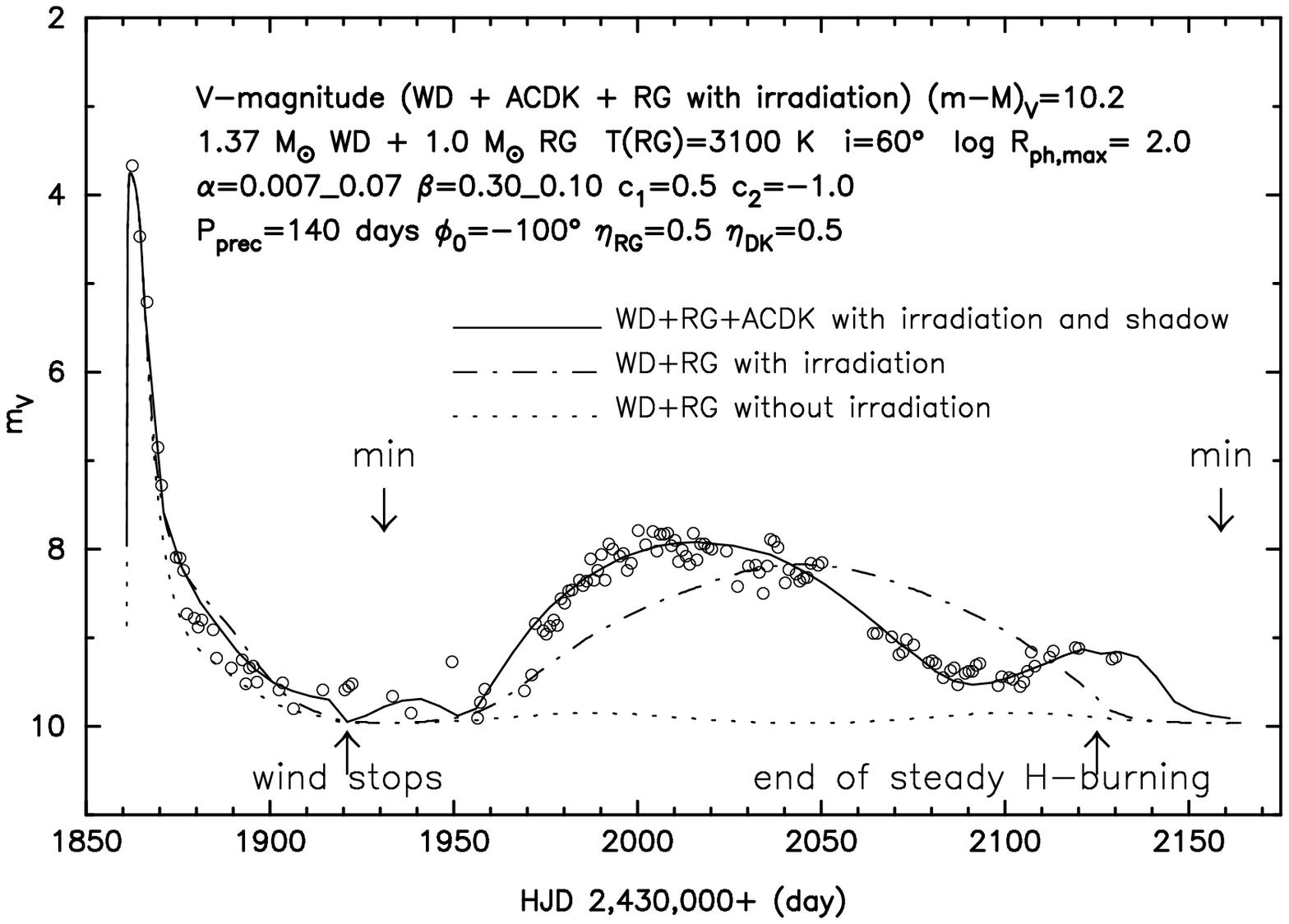}
\caption{(left) Model configuration near the second peak of the 
recurrent nova T CrB.  
The photospheric radius of the hot component at the second peak is 
as small as $\sim 0.003 R_{\odot}$, about $\sim 0.00003$ times the size
of the cool component, and the accretion disk radius  
is $\sim 6 R_{\odot}$, i.e., about $\sim 0.07$ times the
inner critical Roche lobe size.
They are exaggerated so that it may be seen easily.
(light) Light curve fitting of T CrB.
Large arrows attached by "min" indicate epochs at the spectroscopic 
conjunction with the M giant in front.
(Dotted line): The total $V$ light of the white dwarf
photosphere and the red giant (RG) photosphere without irradiation. 
Large arrows attached by "min" indicate epochs at the spectroscopic 
conjunction with the M giant in front. 
(Dash-dotted line): The total $V$ light of the WD photosphere and
the RG photosphere irradiated by the white dwarf.
(Solid line): the total $V$ light of the WD photosphere,
the RG photosphere with irradiation, and the accretion disk surface 
heated-up by the hot component.
The parameters specifying the light curves are shown in the figure.
}
\end{figure}

%%%%%%%%%%%%%%%%   RS Oph  and its figure  %%%%%%%%%%%%%%%%%%%%

Figure 6 shows light curve fitting of RS Oph. In the early phase the 
development of light curve is determined by the wind mass loss, which 
depend only on the white dwarf mass and the heavy element composition 
of the envelope. There are some observational indication that this object 
is an old population, thus we have calculated 
theoretical curves with different metallicities and white dwarf masses.
The left figure shows one of such trial fittings that shows, if $Z=0.004$, 
the best fit curve is obtained with extremely massive white
dwarf $1.377 M_{\odot}$. This is a very critical value just before a
type Ia supernova explosion (Nomoto, Thielemann, \& Yokoi 1984)

The right figure shows the light curve fitting of entire phase of outburst.
In the later phase, the accretion disk and the irradiated companion 
substantially contribute the $V$ light curve and we determine the 
shape and the size of the accretion disk (Hachisu \& Kato 2000a, Hachisu \& 
Kato 2001) . 
The optically thick wind blows during the period from the first
phase of the outburst (HJD 2,446,092) to 79 days after
the maximum (HJD 2,446,170).  The steady hydrogen shell burning
ends at 112 days after maximum (HJD 2,446,203).

%%%%%%%%%%%%%%%%%%%%%%%%%%%%%%%%%%%%%%%%%%%%
\begin{figure}
%\plottwo{kato.rspeak.eps}{kato.rsoph.eps}
\plottwo{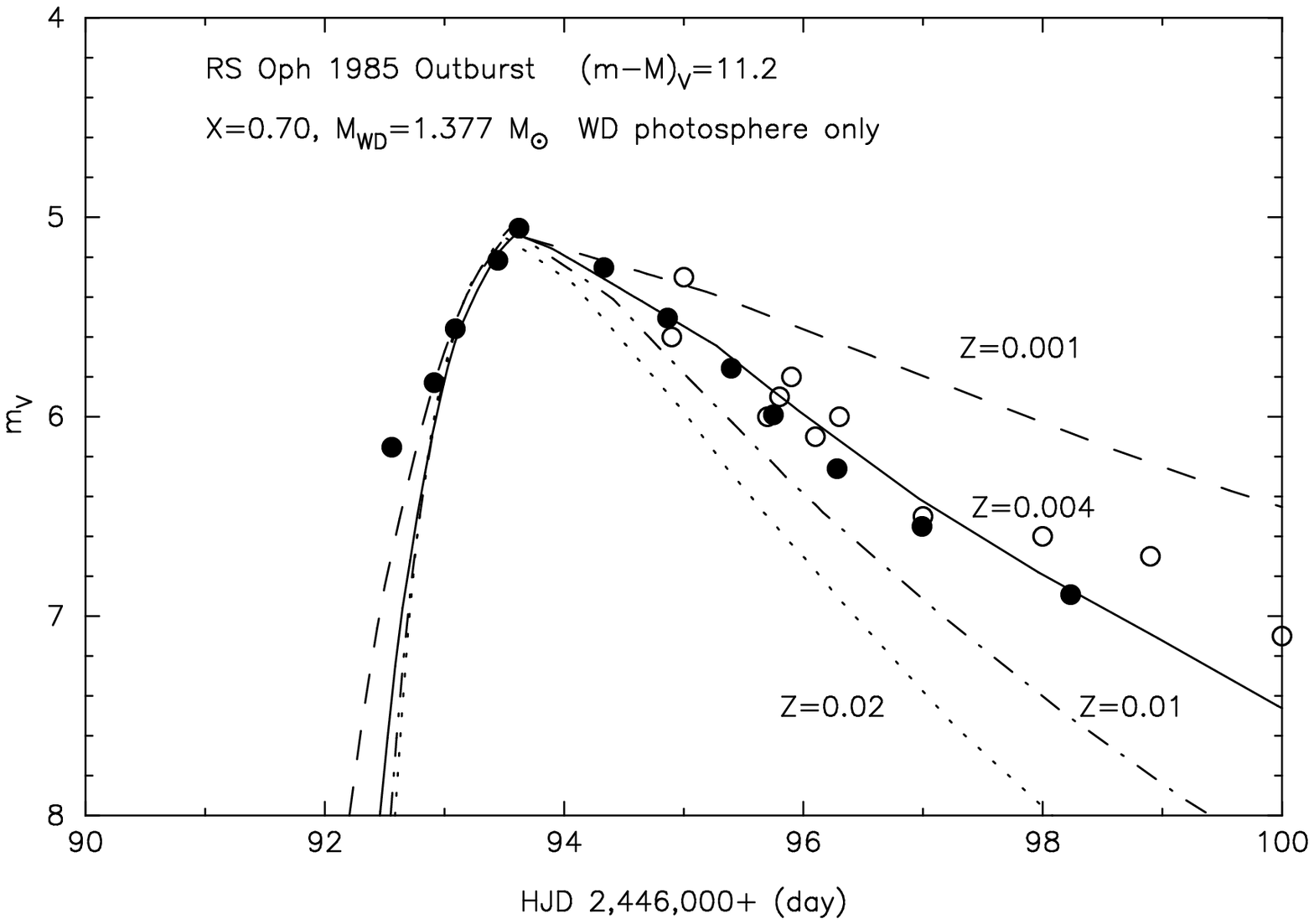}{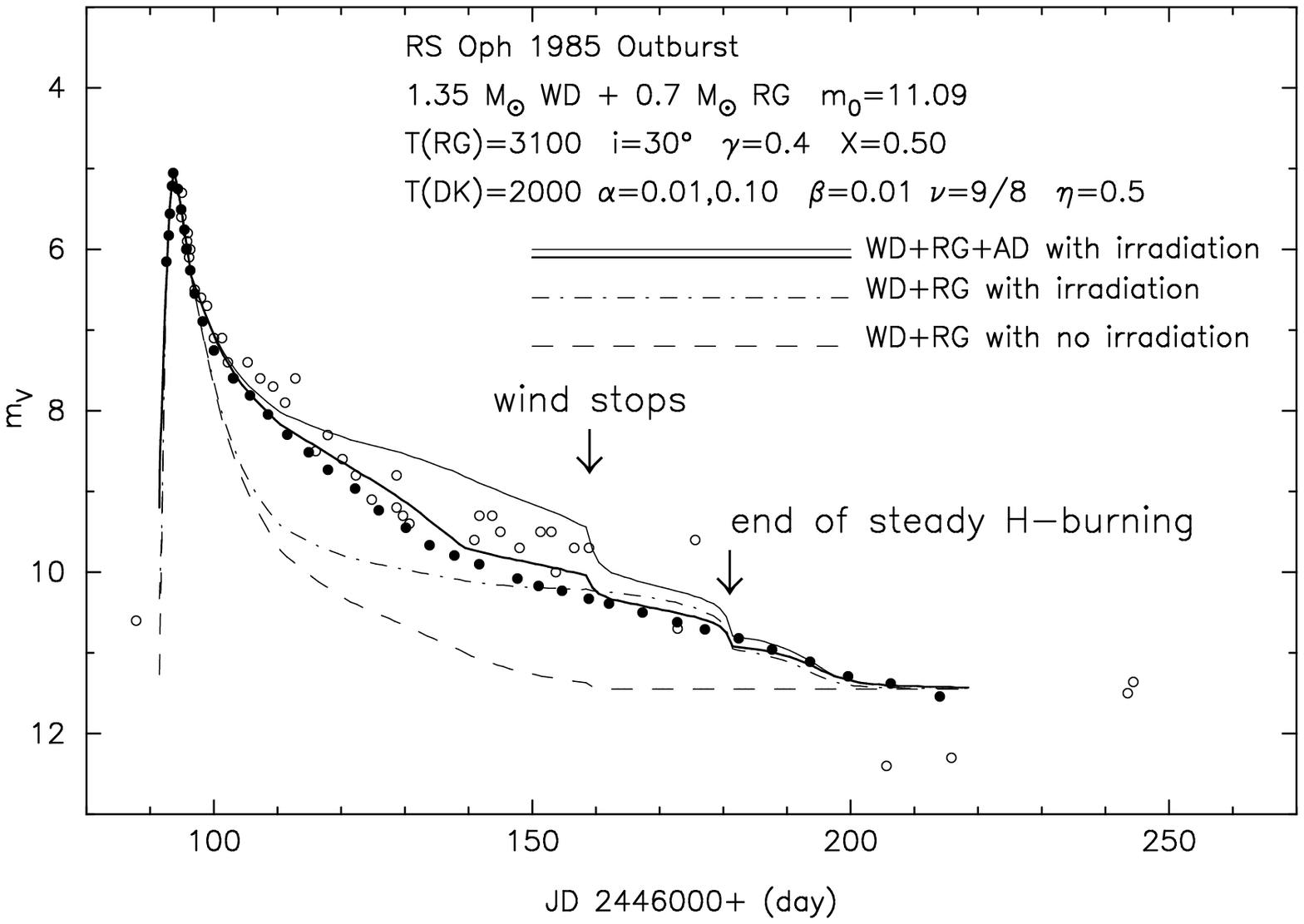}
 \caption{(left) $V$ Light curve fitting of RS Oph in the very early phase. 
The metallicity is attached to each line (for $Z=0.001$, $Z=0.004$,
$Z=0.01$, $Z=0.02$). The hydrogen content of the WD envelope is assumed 
to be $X=0.70$, since the hydrogen content, $X$, hardly affect the 
decline rate of the $V$ light curve (kato 1999).
Filled circles indicate observational points with the previous outbursts. 
%The apparent distance modulus is $(m-M)_V= 11.2$.
(light)
The light curve fitting for the entire outburst phase. We assume
$M_{\rm WD}= 1.377 M_{\odot}$, $X=0.7$, and $Z=0.004$.
(Dashed line): Sum of the contributions from the white dwarf 
photosphere and the nonirradiated red giant photosphere.  
The RG lies well within
the inner critical Roche lobe, i.e., its radius is 0.25 times
the Roche lobe size.
(Dashed-doted line): Sum of the WD photosphere and the irradiated RG 
photosphere.
(Solid line): the total $V$ light of the WD photosphere,
the irradiated RG photosphere and the accretion disk  surface.
The thin and the tick solid curves have different shape of the accretion
disk (for detail see Hachisu and Kato 2001)
}
\end{figure}

%%%%%%%%%%%%%   V745 Sco and V3890 Sgr and their figures (7)   %%%%%%%%%%%
The light curve fitting of the other two objects are shown in figure 7.
The left figure represents the total $V$ light of the WD photosphere,
the irradiated RG, and the irradiated accretion disk.
From the light curve fitting we conclude that the white dwarf mass is as 
massive as $1.35 M_{\odot}$ 
In the same way, the light curve of V3890 Sgr is reproduced as shown in 
the right figure. The white dwarf mass is estimated as  $1.35 M_{\odot}$. 

\begin{figure}
%\plottwo{kato.v745.eps}{kato.v3890.eps}
\plottwo{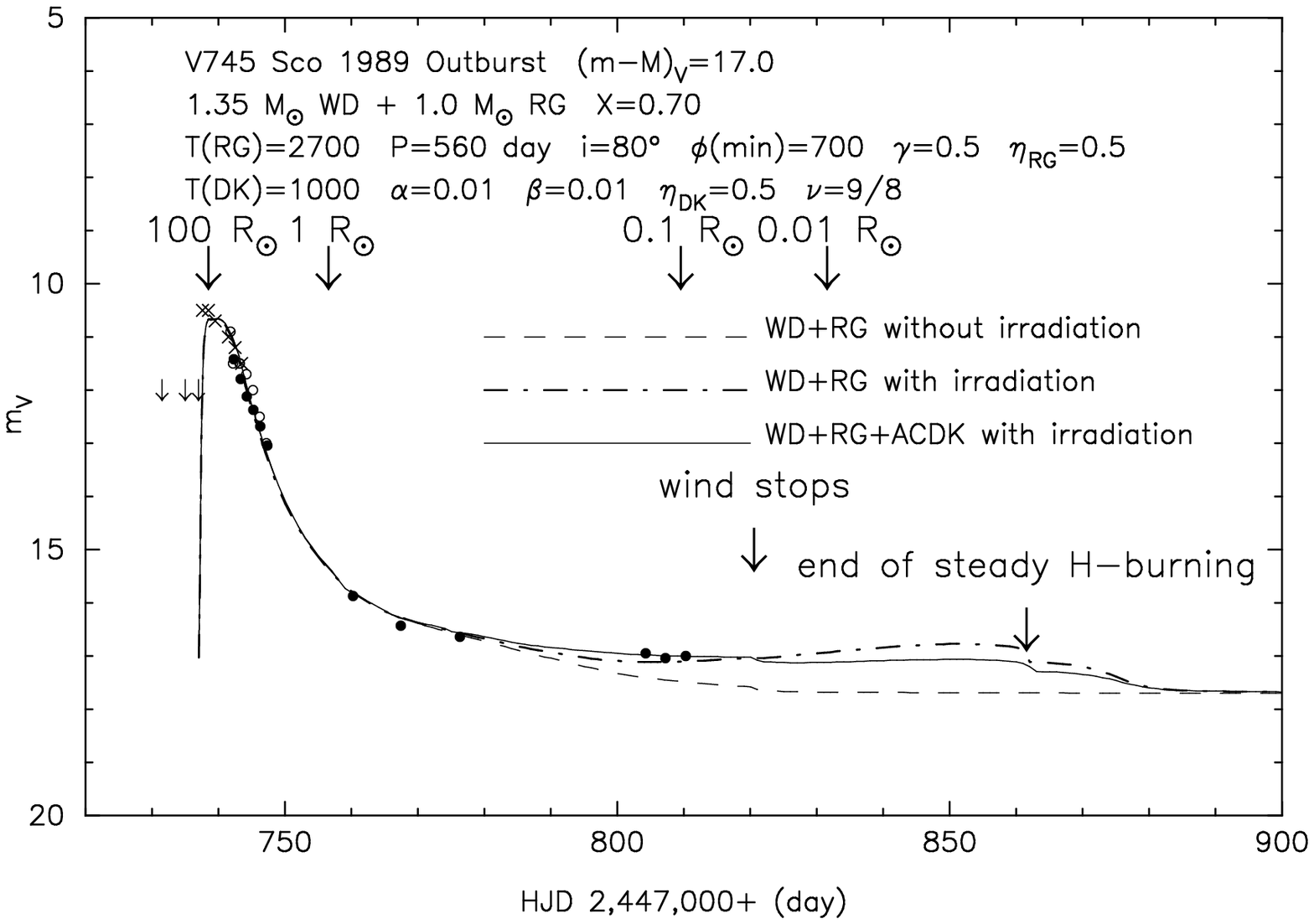}{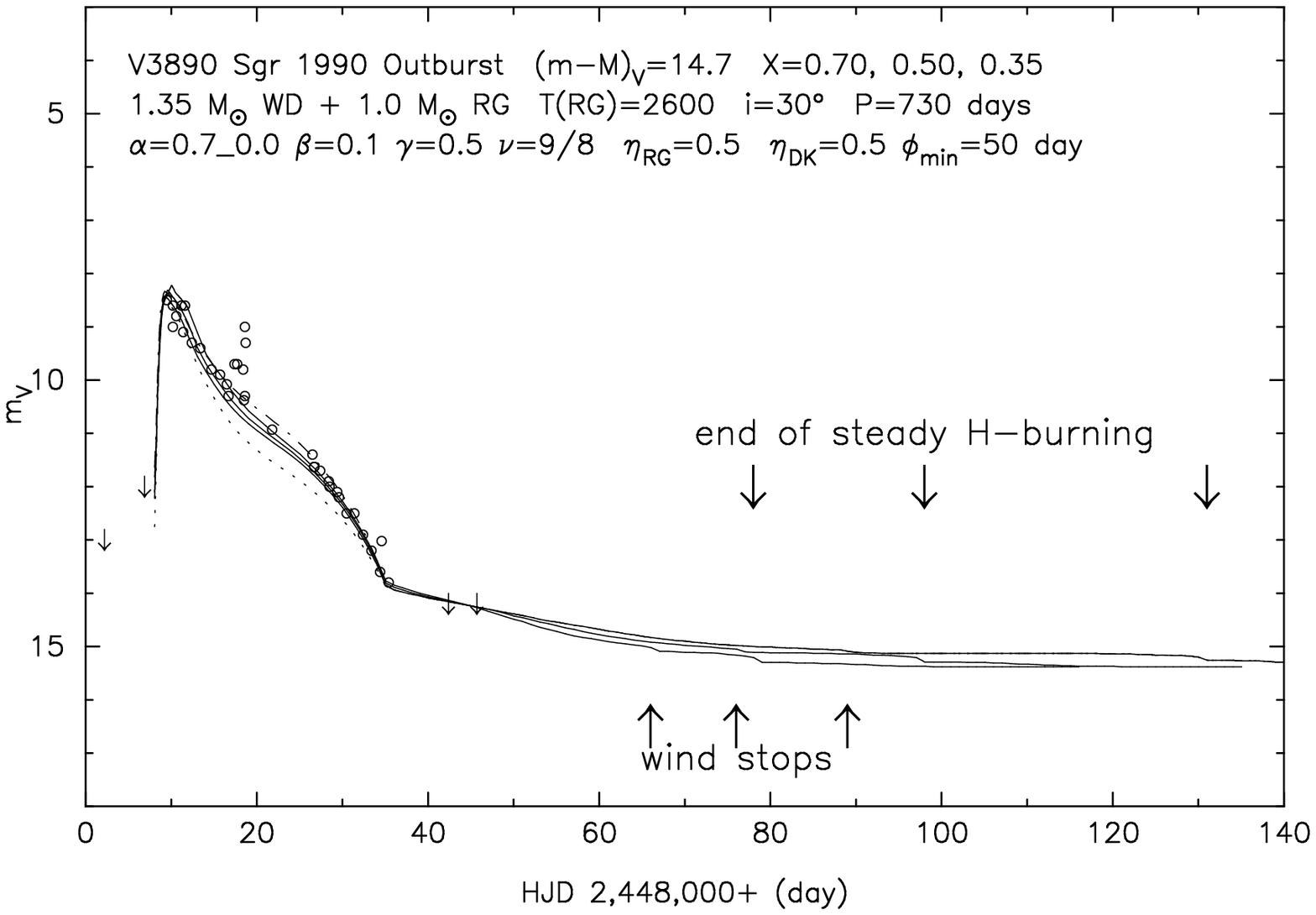}
   \caption{(left) V745 Sco; Model light curves and observational points 
of the  1989 outburst. The solid line represents the best fitted model 
of $1.35 M_{\odot}$ WD, the irradiated RG, and the irradiated accretion disk.
The dashed line denotes the $V$ light curve of the WD photosphere and
the nonirradiated red giant.  The dash-dotted line corresponds to
the $V$ light from the WD photosphere and the irradiated red giant. 
The photospheric radius of the white dwarf is indicated above the light
curve.  The optically thick wind stops at HJD 2,447,819 
and the steady hydrogen shell burning ends at HJD 2,447,860.
(light) V3890 Sgr; Model light curves are plotted against time 
(HJD 2,448,000+). Open circles indicate the observational points taken 
from IAU Circulars of the 1990 outburst.
}
\end{figure}

%%%%%%%%%%%%%%%  candidate of Type Ia SNe
\section{Very Massive White Dwarfs}
Seven out of these eight recurrent novae
contain a very massive white dwarf close to the Chandrasekhar
mass limit  1.35 $M_{\odot}$ for V2487 Oph, V3890 Sgr, and V745 Sco,
1.37 $M_{\odot}$ for T CrB, U Sco, V394 CrA, RS Oph (or more, depending on 
$Z$). Moreover, the growth rate of the white dwarf mass is estimated to be 
$0.1-1 \times 10^{-7}~M_{\odot}yr^{-1}$ (Hachisu \& Kato 2001 and references
therein). Therefore, they are
strong candidates of type Ia SN progenitors.

%%%%%%%%%%%%%%   Last section   %%%%%%%%%%%%%%%%%%%%%%%%%%%%%%%%
\section{Population II novae}
As mentioned in section 1, the acceleration of nova wind in the decay 
phase is governed by the iron peak in the opacity, that sensitively changes 
by heavy element content as shown in figure 1.  
The wind mass-loss is weak in population II novae, because a small iron 
peak produces only weak acceleration. Thus, a systemic differences are
expected between population II novae and disk novae. A population II nova 
will show a smaller decline rate, a longer duration time and slower wind
velocities compared with those of disk nova on the same white dwarf mass
(Kato 1999)

Thus, a statistically trend of slow evolution in light curve is expected 
in population II novae. Fast novae and recurrent novae observed in LMC, 
SMC, and a globular cluster indicate that the binaries in such a low iron
environment have very massive white dwarfs. 
Recent studies of stellar evolution suggest that low metallicity environment 
intends to produce massive white dwarfs (Umeda et al 1999). 

\bigskip

%\acknowledgements{  }

\end{document}